\documentclass[seceq]{ptptex}
\usepackage{bm}
\usepackage{amsmath}
\usepackage{amssymb}
\usepackage[dvips]{graphicx}

\newcommand{\expv}[1]{\langle #1 \rangle}
\newcommand{\tr}{\mbox{tr}}

\newcommand{\pardiff}[2]{\frac{\partial #1 }{\partial #2 }}

\newcommand{\diff}[2]{\frac{d #1}{d #2}}

\newcommand{\sbrace}[1]{\left( #1 \right)}
\newcommand{\mbrace}[1]{\left\{ #1 \right\}}
\newcommand{\bbrace}[1]{\left[ #1 \right]}

\newcommand{\Slash}[1]{\ooalign{\hfil/\hfil\crcr$#1$}}
\usepackage{booktabs}
\newcommand{\PSfig}[2]{\includegraphics[width=#2]{PS/#1}}
\newcommand{\beq}{\begin{eqnarray}}
\newcommand{\eeq}{\end{eqnarray}}
\newcommand{\psibar}{\overline{\psi}}
\newcommand{\btau}{\boldsymbol{\tau}}
\newcommand{\bpi}{\boldsymbol{\pi}}
\newcommand{\brho}{\boldsymbol{\rho}}
\newcommand{\Lag}{{\cal L}}
\newcommand{\VEV}[1]{\langle{#1}\rangle}
\newcommand{\rhoB}{\rho_{\scriptscriptstyle B}}
\newcommand{\rhoBt}{\rho_{{\scriptscriptstyle B}}^\alpha}
\newcommand{\rhoBn}{\rho_{ {\scriptscriptstyle B}}^n}
\newcommand{\rhoBp}{\rho_{ {\scriptscriptstyle B}}^p}
\newcommand{\rhoS}{\rho_{\scriptscriptstyle S}}
\newcommand{\rhoSt}{\rho_{{\scriptscriptstyle S}}^\alpha}
\newcommand{\rhoSn}{\rho_{ {\scriptscriptstyle S}}^n}
\newcommand{\rhoSp}{\rho_{ {\scriptscriptstyle S}}^p}
\newcommand{\rhoT}{\rho_\tau}
\renewcommand{\bold}[1]{\mib{#1}}
\newcommand{\Fint}[1]{{\cal D}[#1]}
\newcommand{\chibar}{\bar{\chi}}
\newcommand{\mathss}[1]{{\scriptscriptstyle \mathrm{#1}}}
\newcommand{\SCL}{{\scriptscriptstyle \mathrm{SCL}}}
\def\nuc#1#2{\relax\ifmmode{}^{#1}{\hbox{#2}}\else${}^{#1}$#2\fi}
\newcommand{\comments}[1]{}

\title{
A chiral symmetric relativistic mean field model \\
with logarithmic sigma potential
}

\author{
Kohsuke \textsc{Tsubakihara} and Akira \textsc{Ohnishi}%
}

\inst{
Department of Physics, Faculty of Science, Hokkaido University, \\
Sapporo 060-0810, Japan
}

\abst{
We develop a chiral symmetric relativistic mean field model
with logarithmic sigma potential 
derived in the strong coupling limit of the lattice QCD.
We find that both of the nuclear matter and finite nuclei
are well described in the present model.
The normal vacuum is found to have global stability at
zero and finite baryon densities,
and an equation of state with moderate stiffness
($K \simeq 280$ MeV) is obtained.
The binding energies and charge radii of $Z$ closed even-even nuclei
are well reproduced in a wide mass range from C to Pb isotopes,
except for the underestimates of binding energies
in several $jj$ closed nuclei.
}

\begin{document}

\maketitle

\section{Introduction}

Chiral symmetry is a fundamental symmetry of QCD at zero quark masses,
and its spontaneous symmetry breaking generates
constituent quark and thus hadron masses.
The Nambu-Goldstone boson of the chiral symmetry,
{\em i.e.} pion, mediates the long range part of nuclear force~\cite{Yukawa},
and the midrange attractive nuclear force can be described
with a light scalar isoscalar meson, so called $\sigma$,
which would be the chiral partner of the pion
representing the fluctuation of the chiral condensate.
Thus it would be desirable to respect the chiral symmetry
in theories of quark, hadron, and nuclear physics.
Actually, many models and theories of quarks and hadrons
such as 
the Nambu-Jona-Lasino model,\cite{NJL,NJL_PRep}
the sigma model,\cite{GellMannLevy_1960}
and the chiral perturbation theories~\cite{CPT}
have been constructed on the basis of the chiral symmetry.

In nuclear many-body problems, 
relativistic mean field (RMF) models have been developed to describe
properties of nuclear matter and finite nuclei.
The first RMF model proposed by Serot and Walecka~\cite{SW86}
involves the coupling of nucleons to the isoscalar vector field,
the omega meson, in addition to the coupling to the sigma meson.
Later on, non-linear self-coupling terms of sigma~\cite{Boguta_phi4,NL1,NL3,TM1}
and omega mesons~\cite{TM1,RBHF} are introduced to obtain better descriptions
of nuclear matter and finite nuclei.
Obtained Lagrangians contain $\sigma^2, \sigma^3, \sigma^4$ terms,
which remind us of the form of the chiral linear sigma model.
RMF models have been successfully applied to various nuclear many-body problems;
the nuclear matter saturation~\cite{SW86}, 
single particle levels in finite nuclei
including the spin-orbit splittings~\cite{SW86},
nuclear binding energies~\cite{NL1,NL3,TM1}
and nucleon-nucleus scattering cross sections and spin observables
~\cite{CHMRS83}.
Furthermore, the equations of state (EOS) constructed with these models
have been successfully applied to compact astrophysical objects
such as neutron stars and supernovae.\cite{Shen,Sumiyoshi,IOS}
Having these successes and the Lagrangian forms in mind,
it would be natural to expect that
RMF is not only a phenomenological model
parameterizing nuclear energy functionals,
but also a starting point of finite baryon density hadronic models
provided that the chiral symmetry is respected.

Contrary to the expectations above,
simple chiral RMF models fail to describe
the nuclei as a many-body system of nucleons.
Since the valence nucleon Fermi integral
prefers smaller chiral condensate,
the normal vacuum jumps to a chiral restored abnormal one at a low density
below the normal nuclear
density in a mean field treatment
of the linear sigma model.\cite{LeeWick_1974,LeeMargulies_1975,Boguta1983}
Thus we need to have a more repulsive potential
than in the linear sigma model at small values of $\sigma$
at vacuum or at finite densities.
There are several prescriptions proposed so far
to generate this repulsive potential 
in order to cure the abnormal vacuum problem at low densities.
One of the prescriptions is to include fermion vacuum
fluctuations~\cite{BL,LeeMargulies_1975,HatsudaPrakash_1998}.
From the baryon one-loop renormalization, 
non-linear and non-analytic sigma potential terms appear and stabilize
the normal vacuum~\cite{BL,LeeMargulies_1975}.
With quark loops, 
one obtains similar terms as in the baryon loops,
and they break the chiral symmetry dynamically.\cite{NJL_PRep}
When one tunes the coefficients 
in the potential terms generated by the quark loops,
it is also possible to prevent the vacuum from falling
into the abnormal vacuum at a small density.\cite{HatsudaPrakash_1998}
In both of the cases, 
a simple interaction term from fermion loops of the form
$-\sigma^4 \log \sigma^2$,
where $\sigma \propto -\langle\bar{q}q\rangle$
represents the light scalar field,
gives rise to instability
at large chiral condensates,
then the normal vacuum is not a real energy minimum state.
This may not be a practical problem in discussing uniform nuclear matter
if we regard the extrema as the normal vacuum,
but in finite nuclei the solution of the scalar meson tends to be anomalous
when the barrier height is low.
In addition, 
we should also include meson loop effects in the baryon loop case,
and meson loops tend to cancel the non-linear sigma terms
coming from the baryon loops~\cite{ref_MesonLoop}
and the normal vacuum would become unstable again at low densities.
The second approach is to introduce the coupling
of sigma and omega mesons.\cite{Boguta1983,Ogawa2004}
Due to the reduction of the omega meson mass according
to the partial chiral symmetry restoration at finite densities,
the repulsive vector interaction becomes strong
then the normal vacuum is kept to be stable even at high densities.
Quantitatively, however, the $\sigma^2\omega^2$ coupling gives rise to
large repulsive vector interaction
and the nuclear matter incompressibility is found to be much larger
than the empirical values.\cite{Boguta1983,Ogawa2004}
In order to soften EOS in this scenario,
we need to include higher order terms
such as $\sigma^6$ and $\sigma^8$~\cite{SO2000}.
In Ref.~\citen{SO2000}, it is demonstrated that we can actually
construct an RMF Lagrangian which gives a soft EOS,
but the coefficients of the higher order potential terms
are negative leading to the instability at large chiral condensates.
The fourth way to overcome the abnormal vacuum problem
is to incorporate the glueball field which simulates the scale anomaly
in QCD.\cite{Heide1994,Furnstahl1993,Furnstahl1995,Mishustin1993,%
Papazoglow1997,Papazoglow1998}
While it may not be justified to include the unobserved glueballs
in hadronic models,
these models give nuclear matter incompressibility
in an acceptable range,
and they roughly explain the bulk properties of nuclei.\cite{Heide1994}
From the symmetry requirement,
the glueball is conjectured to couple with $\sigma$
in the form of $-\chi^4\log\sigma^2$,
where $\chi$ denotes the glueball field.
This potential term ensures the stability of the normal vacuum
in the mean field approximation:
The mass of the glueball is assumed to be heavy, $1-2$ GeV,
then it would be reasonable to consider that
the glueball expectation value is constant (frozen glueball model)
in low energy phenomena.
Under this assumption, the above potential term
is divergent at $\sigma \to 0$ to prevent the normal vacuum from collapsing
to the chirally restored abnormal one.

These approaches are based on QCD inspired effective hadronic or quark models,
and it is preferable to obtain the chiral potential
(energy density as a function of $\sigma$)
directly from QCD.
At present, it is not yet possible to obtain the chiral potential
in Monte-Carlo simulations of the lattice QCD,
since the quark loop contribution is very strong
in the chiral (massless) limit.
One of the promising directions would be to invoke 
the strong coupling limit of the lattice QCD,
where it is possible to perform some analytic evaluations
of the chiral potential in the chiral limit.
Actually, a logarithmic $\sigma$ potential term similar to that
in the glueball model had been already
derived in the strong coupling limit of the lattice QCD.\cite{KS81}
In the strong coupling limit, $1/g^2_{QCD} \to 0$, 
the pure gluonic action disappears and we can perform the one-link
integral analytically.\cite{KS81,Nishida2004,KMOO06%
,SCL-FiniteT,SCL-General}
After the $1/d$ expansion, bosonization,
and fermion integral, we can obtain the effective free energy
with a logarithmic term, $-\log\sigma^2$.\cite{KS81}

In this paper, we study nuclear matter and finite nuclei
in a flavor $\mathrm{SU}(2)$ chirally symmetric
relativistic mean field model
containing a logarithmic potential term of $\sigma$, $-\log\sigma^2$,
derived from the strong coupling limit
of the lattice QCD~\cite{KS81,Nishida2004,KMOO06} at vacuum.
A phenomenological $\omega$ self-interaction term,
$(\omega_\mu\omega^\mu)^2$, is also included in the effective Lagrangian.
Requiring that the pion and nucleon masses
and the pion decay constant are given,
we have four free parameters, $m_\sigma, g_\omega, g_\rho$ and $c_\omega$
($\sigma$ mass, coupling constants of nucleons with $\omega$ and $\rho$ mesons,
and the coefficient of the  $(\omega_\mu\omega^\mu)^2$ term).
Two of these parameters ($g_\omega$ and $c_\omega$)
are determined to fit the nuclear matter saturation point,
$(\rho_0, E/A)=(0.145 \textrm{fm}^{-3}, -16.3 \textrm{MeV})$,
and other parameters ($m_\sigma$ and $g_\rho$) are determined
to reproduce the binding energies of Sn and Pb isotopes.
By choosing these parameters appropriately,
we find that
the obtained EOS is as soft as those in phenomenologically
successful RMF models such as NL and TM models~\cite{NL1,NL3,TM1}.
Bulk properties (binding energies and charge radii)
of proton (sub-)closed even-even nuclei 
are also well explained in a comparable precision
to NL1~\cite{NL1}, NL3~\cite{NL3}, and TM~\cite{TM1} models.

There are several works
including the coupling of nucleons with the negative parity baryons
\cite{HatsudaPrakash_1998,DK89}
and works based on the other chiral partner assignment of pions
(vector manifestations)~\cite{HY01}.
While these are in promising directions
to solve the difficulty in chiral RMF models,
we stick to a naive assignment of nucleon and pion chiral partners
in this work.
Finite temperature and finite chemical potential treatments
in the strong coupling limit of the lattice QCD
would be also necessary to describe the chiral phase transition
at high temperatures and densities,
but these are beyond the scope of this paper.
In order to discuss the properties of nuclear matter and finite nuclei
around their ground states, we expect that the dynamics
in hadronic degrees of freedom dominates.

This paper is organized as follows.
In Sec.~\ref{Sec:Model},
we briefly explain the derivation of the $\sigma$ potential term
in the strong coupling limit of lattice QCD,
and describe our effective hadronic Lagrangian.
In Sec.~\ref{Sec:Results},
we discuss the properties of nuclear matter and finite nuclei
calculated in the present model
in comparison with the results in some other models.
We summarize our work in Sec.~\ref{Sec:Summary}.

\section{Chiral symmetric sigma potential}
\label{Sec:Model}

\subsection{Logarithmic chiral potential
in the strong coupling limit of the lattice QCD}

Chiral potential $V_\sigma$
(energy density as a function of $\sigma$ at vacuum)
is one of the most important ingredients in chiral models;
it dynamically breaks the chiral symmetry,
and the finite expectation value of sigma
generate the constituent quark and hadron masses.
In this paper, we utilize the chiral potential
derived in the strong coupling limit (SCL)
of lattice QCD~\cite{KS81,Nishida2004,KMOO06}.
Here we briefly summarize how to derive
the chiral potential.

In the strong coupling limit ($g \to \infty$),
we can ignore the pure gluonic part
of the lattice QCD action which is proportional to $1/g^2$,
and we keep only those terms including fermions $S_F$ written as,
\beq
S_F[\chi,\chibar,U]
&=&
 \frac12 \sum_{x,\mu}
 	\eta_\mu(x)
	\left[
		 \chibar(x)U_\mu(x)\chi(x+\hat{\mu})
		-\chibar(x+\hat{\mu})U^\dagger_\mu(x)\chi(x)
	\right]
\ ,
\eeq
where $\eta_\mu(x)=(-1)^{x_0+x_1+\cdots+x_{\mu-1}}$, and we express the action
in the lattice unit.
We consider two species of staggered fermions
simulating $u$ and $d$ quarks.
After integrating out the link variable $U_\mu$
in the leading order of $1/d$ expansion
and introducing the auxiliary fields $\sigma_{\alpha\beta}$,
we obtain the following partition function
\beq
{\cal Z} &=& \int \Fint{\chi,\chibar,U}
\exp\left( -S_F[\chi,\chibar,U] - S_U[U] \right)
\nonumber\\
&\simeq& \int\Fint{\chi,\chibar} \exp\left[
	\sum_{x,y,\alpha,\beta}
		{\cal M}_{\alpha\beta}(x)V_M(x,y){\cal M}(y)_{\beta\alpha}
	\right]
\nonumber\\
&=& \int \Fint{\chi,\chibar,\sigma}
	\exp\left(
		-S_\sigma[\chi,\chibar,\sigma]
	\right)
\label{Eq:Partition}
\ ,\\
S_\sigma[\chi,\chibar,\sigma] &=&
		\sum_{x,y,\alpha,\beta}
		\sigma(x)_{\alpha\beta}V_M^{-1}(x,y)\sigma(y)_{\beta\alpha}
		+2\sum_{x,\alpha,\beta}
			\sigma(x)_{\alpha\beta}{\cal M}(x)_{\beta\alpha}
\ .
\label{auxiliary}
\eeq
Mesonic composites are defined as
${\cal M}_{\alpha\beta}(x)=\chibar^a_\alpha(x)\chi^a_\beta(x)$,
where the superscript $a$ denotes color
and the subscripts $\alpha$ and $\beta$ show the flavor.
The lattice mesonic inverse propagator $V_M(x,y)$ is given as
$V_M(x,y)=\sum_\mu
\left(\delta_{y,x+\hat{\mu}} + \delta_{y,x-\hat{\mu}}\right)/8N_c$.
From the first to the second line in Eq.~(\ref{Eq:Partition}),
we have used the one-link integral formula,
$\int dU U_{ab}U^\dagger_{cd}=\delta_{ad}\delta_{bc}/N_c$.
The auxiliary fields are related to the expectation values
of the mesonic composites as
$\VEV{\sigma_{\alpha\beta}(x)}=-\VEV{V_M(x,y){\cal M}_{\alpha\beta}(y)}$.

Now we consider static and uniform scalar $\sigma$
and pseudoscalar $\bpi$ condensates,
and we substitute the auxiliary fields by the mean field ansatz,
\beq
\sigma_{\alpha\beta}(x)=
\begin{pmatrix}
\frac{1}{\sqrt{2}}(\sigma+i\varepsilon(x)\pi^0) & i\varepsilon(x)\pi_c^* \\
i\varepsilon(x)\pi_c & \frac{1}{\sqrt{2}}(\sigma-i\varepsilon(x)\pi^0)
\end{pmatrix}\ ,
\eeq
where $\varepsilon(x)=(-1)^{x_0+x_1+x_2+x_3}$.
Since fermions are decoupled in each space-time point,
we can easily perform the fermion integral.
The effective free energy is obtained as,
\beq
V_\chi(\sigma,\bpi) &=& \VEV{\tr \left[\sigma V_M^{-1} \sigma\right]}
- N_c\log\det{\sigma_{\alpha\beta}}
= \VEV{V_M^{-1}}\mbox{tr}\left[M^\dagger M\right]
- N_c\log\det{\sigma_{\alpha\beta}}
\nonumber\\
&=& N_c\phi^2 - N_c \log\phi^2
\quad
(\phi^2 = \sigma^2+\bpi^2)
\ ,
\label{Eq:Vchi_SCL_A}
\eeq
where $\VEV{\cdots}$ denotes the space-time average,
and $M$ represents the meson matrix
in which the $\varepsilon(x)$ factor is omitted from $\sigma_{\alpha\beta}$,
\beq
M=
\begin{pmatrix}
\frac{1}{\sqrt{2}}(\sigma+i\pi^0) & i\pi_c^* \\
i\pi_c & \frac{1}{\sqrt{2}}(\sigma-i\pi^0)
\end{pmatrix}\ .
\eeq

While the coefficients of $\phi^2$ and $-\log\phi^2$ are given
in Eq.~(\ref{Eq:Vchi_SCL_A}),
these coefficients are fragile.
In addition to the freedom in the choice of the lattice spacing
and the scaling factor for $\sigma$ and $\bpi$ to be in the canonical form,
it has been shown that
the baryonic composite contribution modifies the coefficient
of $\phi^2$~\cite{KMOO06}.
Furthermore, two species of staggered fermions corresponds to $N_f=8$, 
and the coefficient modification may not be trivial when we take $N_f=2$.
Thus we regard them as parameters to obtain physical masses
of $\sigma$ and $\pi$ mesons,
and adopt the following coefficients in the later discussions.
\beq
V_\sigma^\SCL = V_\chi(\sigma,\bpi)-c_\sigma\,\sigma
&=& b_\sigma\phi^2 - a_\sigma\log\phi^2 -c_\sigma\,\sigma
\label{Eq:VsigmaSCL}
\ ,\\
a_\sigma  =  {f_\pi^2 \over 4}(m_\sigma^2 - m_\pi^2)\ ,\quad
b_\sigma &=& {1 \over 4}(m_\sigma^2 + m_\pi^2)\ ,\quad
c_\sigma  =  f_\pi m_\pi^2\ .
\eeq
After requiring that $V_\sigma$ has a minimum at $\sigma=f_\pi$
and fitting the pion mass $m_\pi$, one parameter $m_\sigma$ is left
as a free parameter.

Because of the singularity of $V_\sigma$
at $\sigma \to 0$ which comes from the logarithmic term,
chiral symmetry restoration is suppressed with this chiral potential.
One can doubt that this singularity may come from an artifact
of the strong coupling limit.
Indeed, in a finite temperature treatment of the strong coupling limit,
where the anti-periodic boundary condition of fermions are taken care of,
we do not have a divergent behavior at $\sigma \to 0$,\cite{SCL-FiniteT}
but we still have a finite negative derivative at $\sigma = 0$
even in the chiral limit when $T=0$.
This finite negative derivative at $\sigma=0$ is enough to suppress
the full chiral restoration at finite density,
because the nucleon Fermi integral contribution behaves as $\rhoB\sigma^2$
and we always have a minimum at a finite $\sigma$ value.
Therefore, we consider that the present chiral potential $V_\sigma$
would be a good starting point to describe cold nuclear matter and nuclei.
For finite temperatures, the singularity disappears also in the derivative,
and the full chiral restoration will take place.
In that case, we have to take account of the finite temperature effects
in $V_\sigma$.\cite{Nishida2004,KMOO06}.

\subsection{Comparison with other models}

There is a variety of possibilities
for the choice of the chiral potential $V_\sigma(\phi^2)$.
%
The simplest one is that in the chiral linear $\sigma$ model
($\phi^4$ theory) proposed by Gell-Mann and Levy~\cite{GellMannLevy_1960}.
\beq
V_\sigma^{(\phi^4)}
={\lambda\over4}(\phi^2-f_\pi^2)^2 +\frac12m_\pi^2\phi^2 -f_\pi m_\pi^2\sigma
\ ,\quad
\lambda = {m_\sigma^2-m_\pi^2 \over 2f_\pi^2}
\ .
\eeq
With this chiral potential,
we have only one free parameter $m_\sigma$
and the theory is renormalizable.
%
The double-well structure in the linear $\sigma$ model
can be understood as a result of quark-antiquark condensation
dynamically generated from the chiral symmetric
four Fermi interaction of quarks
in the Nambu-Jona-Lasino(NJL) model.\cite{NJL,NJL_PRep}.
The chiral potential in NJL is found to be
\begin{gather}
V_\sigma^\mathss{NJL}
={m_0^2\over2}\sigma^2 + \Lambda^4 f_\mathss{NJL}\left({G\sigma\over\Lambda}\right)-f_\pi m_\pi^2\sigma
\ ,\\
f_\mathss{NJL}(x)
=-{N_c N_f \over 4\pi^2} \left[
		\left(1+{x^2\over2}\right)\sqrt{1+x^2}-1-{x^4\over 2}\,\log\left({1+\sqrt{1+x^2}\over x}\right)
	\right]
\ ,
\end{gather}
where $G$ represents the coupling of the quark to $\sigma$
and $\pi$ mesons.\cite{NJL_PRep}
In Ref.~\citen{NJL_PRep},
parameter values are fixed to fit $f_\pi$ and $m_\pi$,
resulting in the constituent quark mass $M_q=G f_\pi = 335$ MeV
and cut off $\Lambda=631$ MeV.

%
Nucleon one-loop renormalization of the chiral linear $\sigma$ model
leads to additional potential terms at vacuum.
In the chiral limit ($m_\pi=0$),
the following interaction appears,\cite{BL}
\begin{gather}
V_\sigma^\mathss{BL}
={m_\sigma^2\over 8f_\pi^2}(\phi^2-f_\pi^2)^2
-M_N^4 f_\mathss{BL}(\phi/f_\pi)
\ ,\\
f_\mathss{BL}(x)
=-{1 \over 4\pi^2} \left[
		{x^4\over 2}\log x^2 - \frac14 + x^2 - \frac34 x^4
	\right]
\ .
\end{gather}
In Ref.~\citen{BL}, the value of sigma mass is 
taken to be $m_\sigma=572.8$ MeV.

\begin{figure}[tb]
\centerline{\PSfig{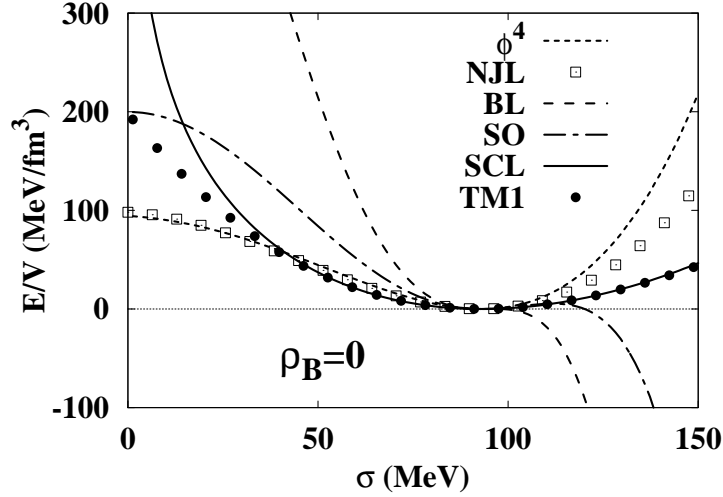}{10cm}}
\caption{
Energy density at vacuum.
Energy densities at vacuum as a function of $\sigma$ 
calculated in the SCL model (solid curve)
are compared with those in the
linear $\sigma$ ($\phi^4$, dotted curve),
NJL~\protect{\cite{NJL,NJL_PRep}} (open squares),
baryon loop (BL, dashed lines)~\protect{\cite{BL}},
Sahu-Ohnishi (SO, dot-dashed lines)~\protect{\cite{SO2000}},
and TM1 (filled circles)~\protect{\cite{TM1}} models.
}
\label{Fig:Vac}
\end{figure}

In Ref.~\citen{SO2000}, Sahu and Ohnishi proposed to add
higher order terms to obtain reasonable incompressibility
within the Boguta scenario~\cite{Boguta1983},
and the chiral potential is given in the chiral limit as follows,
\begin{gather}
V_\sigma^\mathss{SO}={m_\sigma^2\over 8f_\pi^2}(\phi^2-f_\pi^2)^2
+f_\pi^4 f_\mathss{SO}(\phi/f_\pi)
\ ,\\
f_\mathss{SO}(x)={C_6\over 6}(x^2-1)^3 + {C_8\over 8}(x^2-1)^4
\ ,
\end{gather}
with parameters $m_\sigma=762.3$ MeV, 
$C_6 = -74.4$ and $C_8=-2.2$.

For comparison, we also refer here the potential
in a non-chiral model, TM1~\cite{TM1}.
\begin{gather}
V_\sigma^\mathss{TM1}(\varphi)
={m_\sigma^2\over 2}\varphi^2 + {g_3 f_\pi \over 3}\varphi^3 + {g_4\over 4}\varphi^4
\ ,
\end{gather}
where $\varphi$ stands for the deviation from the vacuum,
$\varphi = f_\pi - \sigma$.
Parameters in the TM1 model are given as
$m_\sigma=511.198$ MeV, $g_3 = 15.3383$ and $g_4=0.6183$.

In Fig.~\ref{Fig:Vac},
we compare the chiral potential in the present SCL model,
$V_\sigma^\mathss{SCL}$ in Eq.~(\ref{Eq:VsigmaSCL}),
with those in other models.
The chiral potential in the NJL model (open squares)
describes that in the linear $\sigma$ model ($\phi^4$, dotted line)
in the region $\sigma < f_\pi$,
giving the foundation of the $\phi^4$ model from quark degrees of freedom.
As mentioned in the introduction,
the nucleon Fermi contribution prefers smaller $\sigma$
values at finite density,
and in order to keep $\sigma$ from collapsing to the abnormal
vacuum ($\sigma \sim 0$) at low densities,
we have to have more repulsive potential in the region $\sigma < f_\pi$
than in the linear $\sigma$ model.
While the baryon loop (BL, dashed line)
and Sahu-Ohnishi (SO, dash-dotted line) models
have potentials repulsive enough at small $\sigma$ values,
these models are found to have instability at large $\sigma$ values.
In the present model (SCL, solid line),
we do not have any instability at any values of $\sigma$,
and we have stronger repulsion in the region of $\sigma < f_\pi$.
It is interesting to find that
the SCL model results are very close to that in the TM1 (filled circles)
except for the diverging behavior at $\sigma \to 0$.

\section{Nuclear matter and finite nuclei in chiral RMF models}
\label{Sec:Results}

Relativistic Mean Field (RMF) approach has been developed 
as an effective theory to describe nuclear matter and finite nuclei
in a field theoretical treatment
containing scalar and vector meson couplings to nucleons.
In this paper, we consider the following chiral symmetric RMF Lagrangian
in which nucleons couple with 
$\sigma$, $\bpi$, $\omega$ and $\brho$ fields,
\begin{eqnarray}
\Lag_\chi &=& \psibar_N \bigl[
	i\Slash\partial - g_\sigma(\sigma+i\gamma_5\btau\cdot\bpi)
		-g_\omega\Slash\omega-g_\rho\btau\cdot\Slash\brho
		\bigr] \psi_N
\nonumber\\
	&+& \frac12\left(
		\partial^\mu\sigma\partial_\mu\sigma
		+\partial^\mu\boldsymbol{\pi}\cdot\partial_\mu\boldsymbol{\pi}
		\right)
	- V_\sigma(\sigma,\bpi)
\nonumber\\
	&-& \frac14 W^{\mu\nu}W_{\mu\nu}
		+ \frac12 m_\omega^2 \omega^\mu\omega_\mu
		+ {c_\omega\over 4} (\omega^\mu\omega_\mu)^2
	- \frac14 \bold{R}^{\mu\nu}\cdot\bold{R}_{\mu\nu}
		+ \frac12 m_\rho^2 \brho^\mu\cdot\brho_\mu
\ ,\nonumber\\\label{Eq:Lagrangian}\\
W_{\mu\nu} &=& \partial_\mu\omega_\nu - \partial_\nu \omega_\mu
\ ,\\
\bold{R}_{\mu\nu} &=& \partial_\mu\brho_\nu - \partial_\nu \brho_\mu
		+ g_\rho \brho_\mu \times \brho_\nu
\ .
\end{eqnarray}
Here we have omitted the photon field,
which we include for finite nuclear study.

In this section, we study uniform nuclear matter and finite nuclei
with the Lagrangian in Eq.~(\ref{Eq:Lagrangian})
together with the logarithmic chiral potential $V_\sigma^\SCL$
in Eq.~(\ref{Eq:VsigmaSCL}).
We search for an appropriate parameter set, containing 
the $\sigma$ mass ($m_\sigma$),
meson-nucleon coupling constants ($g_\omega$ and $g_\rho$),
and the strength of the $\omega$ self-interaction ($c_\omega$),
which explains the properties of symmetric nuclear matter and finite nuclei.

\subsection{Nuclear Matter}
\label{Subsec:Matter}

First, we study the EOS of uniform symmetric nuclear matter.
We assume that the meson fields are static and uniform,
then the RMF Lagrangian for nuclear matter becomes
\begin{align}
\Lag_\chi^{\scriptscriptstyle \mathrm{Unif}}
&=\psibar_N\left(i\Slash\partial
-g_\sigma\sigma-\gamma^0(g_\omega\omega+g_\rho\tau_3 R)\right)\psi_N
\nonumber\\
& + \frac{1}{2}m_\omega^2 \omega^2 + \frac{c_\omega}{4} \omega^4
  + \frac{1}{2}m_\rho^2 R^2
  - V_\sigma(\sigma)
\ ,
\end{align}
which includes $\sigma$, $\omega$ and $\rho$ (written as $R$)
mesons.
Here we have omitted the Lorentz and isospin indices,
$\omega = \omega_0$ and $R = \rho_{30}$, for simplicity.

The energy density in symmetric nuclear matter
as a function of $\sigma$ and $\omega$ can be written as,
\begin{equation}
E/V = g_N \int \frac{d\bold{p}}{(2\pi)^3} \sqrt{p^2 + {M_N^\ast}^2 (\sigma)}
          + g_\omega\omega\rhoB - \frac{m_\omega^2}{2}\omega^2 
          - \frac{c_\omega}{4} \omega^4
	  + V_\sigma(\sigma)
\ ,
\label{epv}
\end{equation}
where $g_N=4$ is the spin-isospin degeneracy factor of nucleons,
and $M_N^\ast(\sigma)=g_\sigma \sigma$ is the nucleon effective mass.
The equilibrium conditions read
\begin{gather}
\pardiff{(E/V)}{\sigma} =
 g_\sigma \rhoS + \pardiff{V_\sigma}{\sigma}= 0
\ ,\\
\pardiff{(E/V)}{\omega}
= g_\omega\rhoB - m_\omega^2\omega - c_\omega\omega^3 = 0
\ ,\\
\rhoS = g_N \int \frac{d\bold{p}}{(2\pi)^3}
	{M_N^\ast \over \sqrt{p^2 + {m_N^\ast}^2 (\sigma)}}
\ .
\end{gather}

\begin{figure}[tb]
\centering
\PSfig{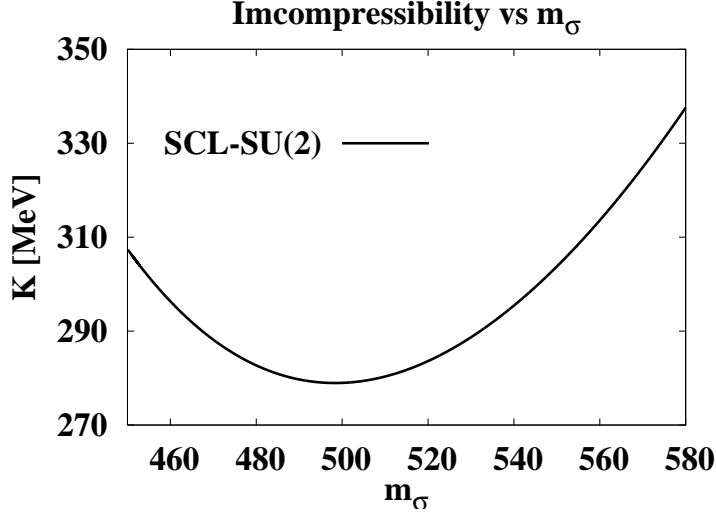}{10cm,clip}
\caption{
Symmetric nuclear matter imcomplessibility $K$.
The symmetric nuclear matter incompressibility $K$
as a function of $\sigma$ mass ($m_\sigma$) in the SCL model is shown.
For a given $m_\sigma$ value,
the $\omega$-nucleon coupling constant ($g_\omega$)
and the $\omega$ self-interaction strength ($c_\omega$) are 
determined to fit the symmetric nuclear matter saturation
point, $(\rho_0, E/A)$,
and the incompressibility $K$ is obtained as a result of fitting.
}
\label{KperMsig}
\end{figure}
In symmetric nuclear matter, we have three relevant parameters,
$m_\sigma, g_\omega$ and $c_\omega$.
When we give $m_\sigma$ as a free parameter,
then other two are determined to fit the saturation properties,
$(\rho_0, E/V)= (0.145~\mathrm{fm}^{-3}, -16.3~\mathrm{MeV})$.
In Fig.~\ref{KperMsig}, we show the nuclear matter incompressibility
$K=9\rho_0^2(\partial^2(E/V)/\partial \rhoB^2)$
as a function of $m_\sigma$.
We find that the incompressibility is smaller than 300 MeV
in the mass region
$460~\textrm{MeV} \lesssim m_\sigma \lesssim 540~\textrm{MeV}$,
which can be regarded as the allowed region.
Especially,
at around the incompressibility minimum
($K \simeq 279~\textrm{MeV}$ at $m_\sigma \sim 500$ MeV),
we obtain EOS which is as soft as 
that in the TM1~\cite{TM1} model.

\begin{figure}[tb]
\centerline{\PSfig{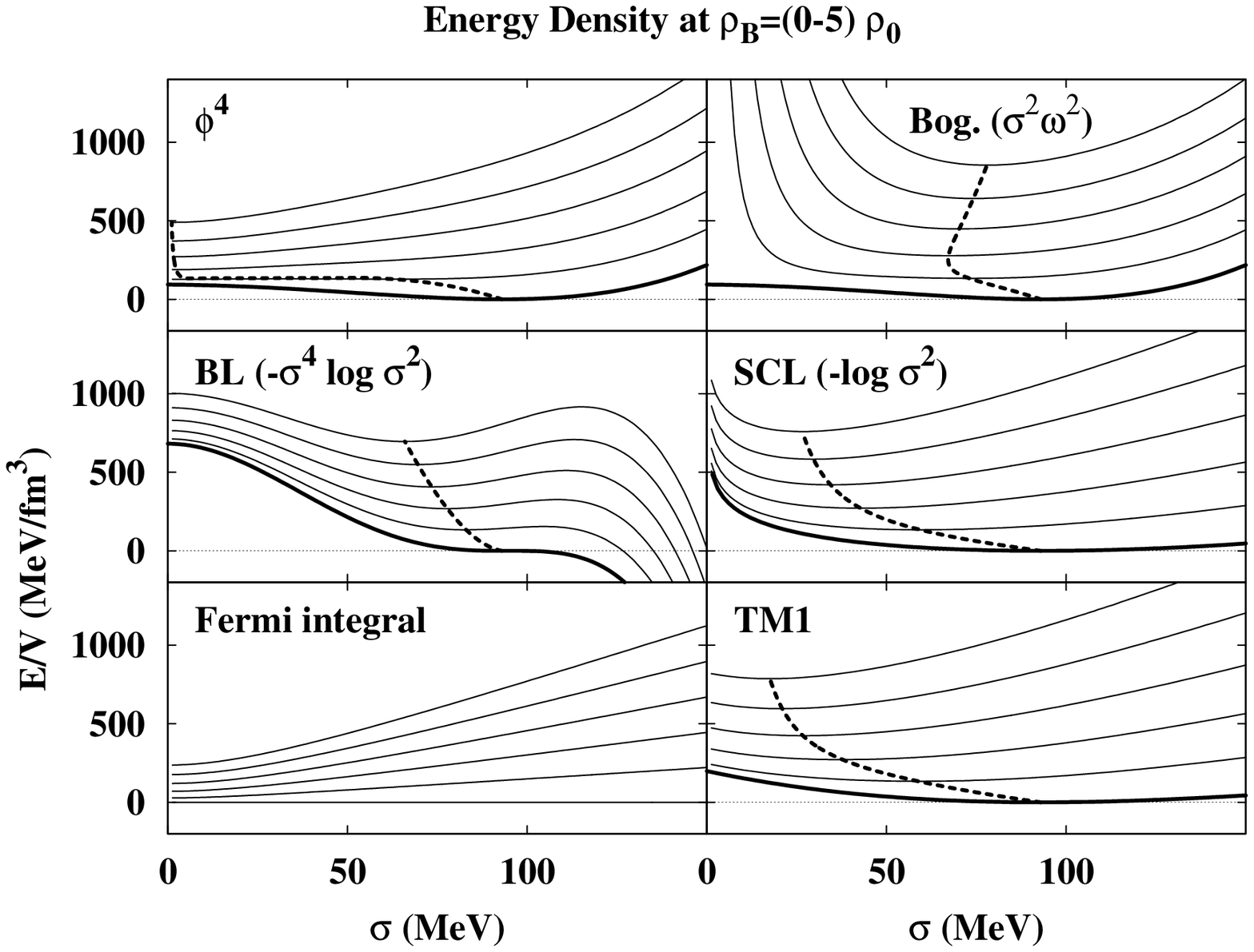}{12cm}}
\caption{
Energy density at finite densities.
Calculated energy densities at vacuum (thick solid curves)
and $\rhoB = (1-5) \rho_0$
(thin solid curves, from down to up as the density grows)
as a function of $\sigma$ 
are compared in chiral RMF models;
the linear $\sigma$ model ($\phi^4$, top left panel),
the Boguta model (Bog., top right panel)~\protect{\cite{Boguta1983,Ogawa2004}},
the baryon loop model (BL, middle left panel)~\protect{\cite{BL}},
and the SCL model (middle right panel).
The nucleon Fermi integral contribution (bottom left panel)
and the results in the TM1 model (bottom right panel)~\protect{\cite{TM1}}
are also shown.
The dotted curves show the equilibrium point where the energy density
becomes the local minimum in the range of $\sigma < f_\pi$.
}
\label{Fig:CompVsig}
\end{figure}

In Fig.~\ref{Fig:CompVsig},
we compare the energy density as a function of $\sigma$
at $\rhoB=(0-5)\rho_0$ in several chiral RMF models.
In the $\phi^4$ model (left top panel),
the nucleon Fermi integral contribution
(left bottom panel in Fig.~\ref{Fig:CompVsig})
is stronger than the repulsive potential at $\sigma < f_\pi$,
and the vacuum collapses to the abnormal one
at a density below
$\rho_0$.\cite{LeeWick_1974,LeeMargulies_1975,Boguta1983,Ogawa2004}
In order to avoid this collapsing,
Boguta~\cite{Boguta1983} replaced the vector meson mass term
$m_\omega^2\omega^2/2$ with that of the $\sigma^2\omega^2$ coupling term,
\begin{equation}
\Lag_{\sigma\omega}^{\scriptscriptstyle \mathrm{Boguta}}
= \frac12 {m_\omega^2 \over f_\pi^2}\,\sigma^2\omega^2
\ .
\end{equation}
In the case of no $\omega$ self-interactions, $c_\omega=0$,
the above coupling term gives large $\omega$ values at small $\sigma$ as,
$\omega=f_\pi^2 g_\omega \rhoB/m_\omega^2\sigma^2$,
leading to a large repulsive potential at finite densities.
Because of this repulsion,
nuclear matter EOS becomes stiff ($K > 600$ MeV),
and $\sigma$ increases again at around $\rhoB \simeq 0.27~\mbox{fm}^{-3}$,
as shown in the top right panel of Fig. \ref{Fig:CompVsig}.

\begin{figure}[tbh]
\centerline{\PSfig{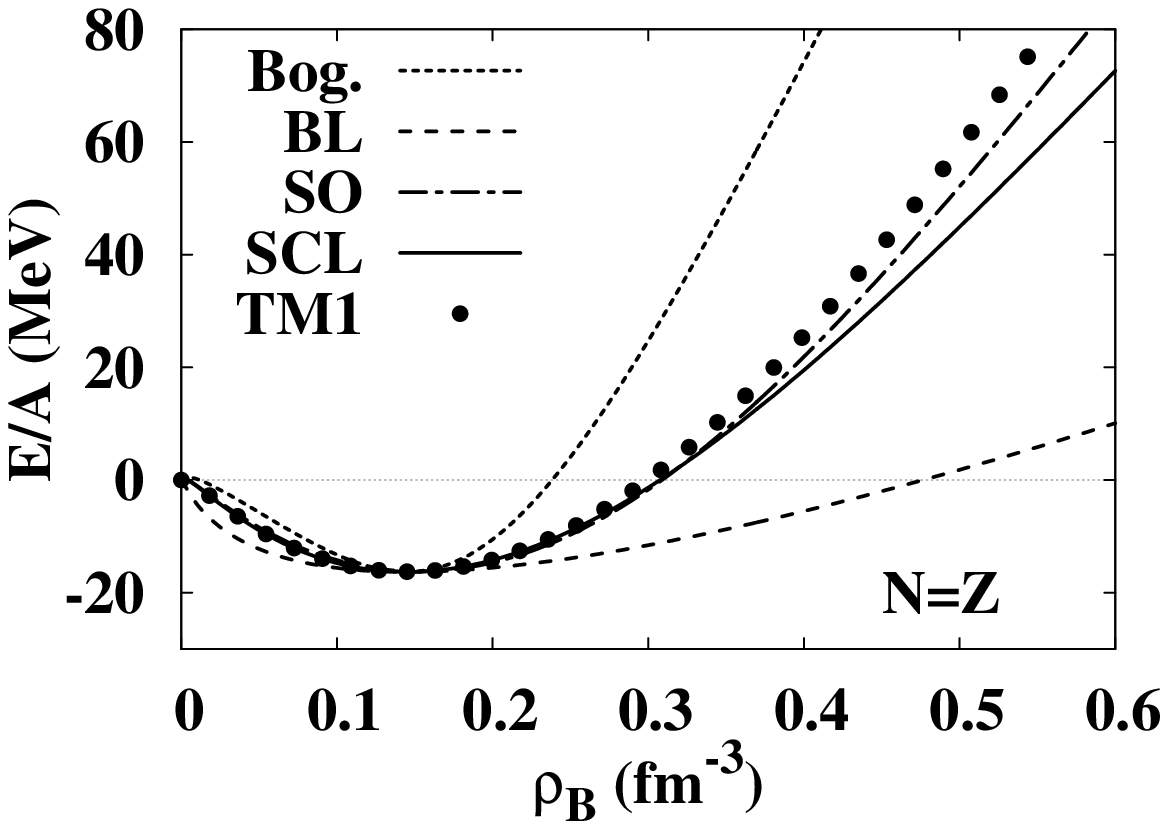}{7.5cm}\PSfig{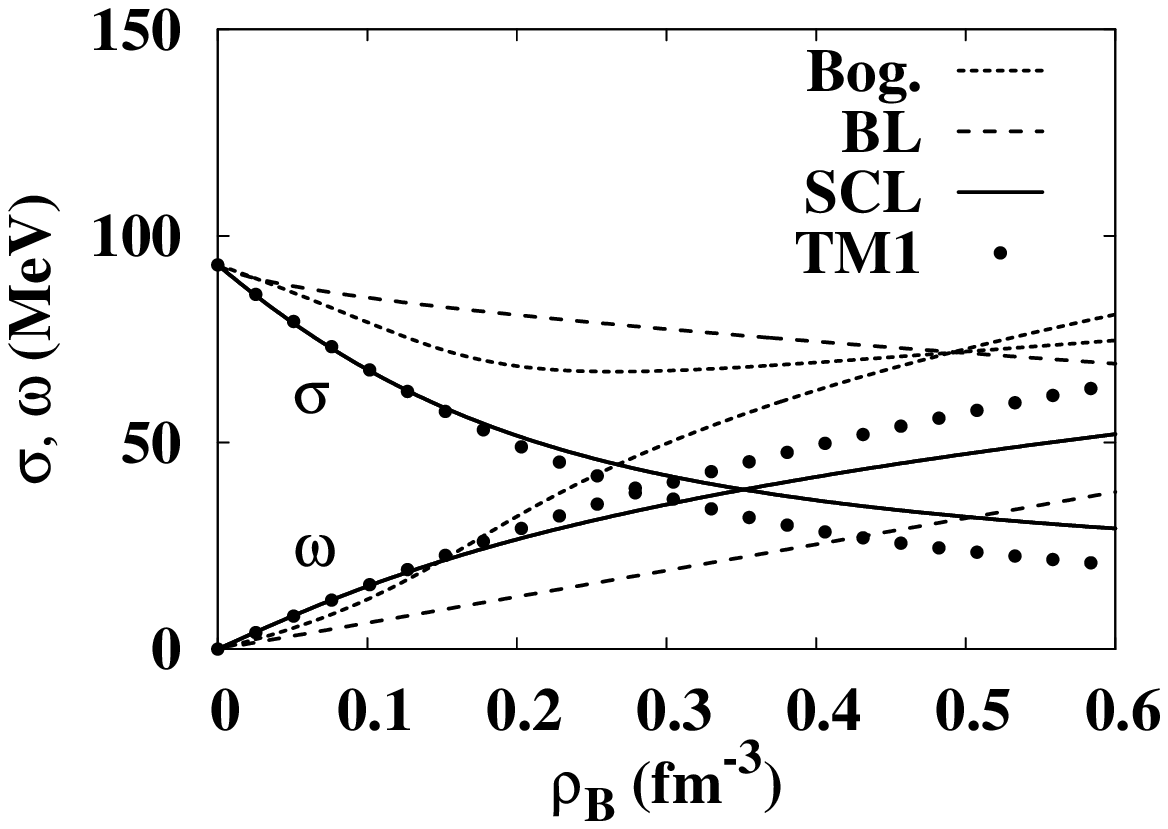}{7.5cm}}
\caption{
Equation of state of symmetric nuclear matter (left panel) and 
density dependence of meson fields (right panel).
We show
the binding energy per nucleon in symmetric nuclear matter (left)
and the $\sigma$ and $\omega$ field expectation values (right)
as a function of the baryon density.
We compare the calculated results in SCL (solid curves)
with those in the
Boguta (Bog., dotted curve)~\protect{\cite{Boguta1983,Ogawa2004}},
baryon loop (BL, dashed lines)~\protect{\cite{BL}},
Sahu-Ohnishi (SO, dot-dashed lines)~\protect{\cite{SO2000}},
and TM1 (filled circles)~\protect{\cite{TM1}} models.
}\label{Fig:EOSes-UsvComp}
\end{figure}

In the left panel of Fig.~\ref{Fig:EOSes-UsvComp},
we compare the EOS in the present model
with those in other chiral RMF models and the TM1 model.
We adopt $m_\sigma = 502.63$ MeV as a typical value,
which gives $K=279.14$ MeV.
Other parameters are summarized in Table~\ref{Table:paramset}.
We find that the present model (SCL) gives softer EOS
than that in the linear sigma model in the Boguta scenario
(Bog.),\cite{Boguta1983,Ogawa2004}
and its incompressibility is comparable to those
in Sahu-Ohnishi (SO)~\cite{SO2000} and TM1~\cite{TM1} models.
The baryon loop (BL) model~\cite{BL} gives
the softest EOS in the models considered here.
However, the BL model has the instability at large $\sigma$ values,
and the incompressibility is too small, $K \simeq 110$ MeV.

In addition to the similarity in EOS,
$\sigma$ and $\omega$ expectation values in the SCL model 
behave in a similar way to those in the TM1 model at low densities.
In the right panel of Fig.~\ref{Fig:EOSes-UsvComp},
we show $\sigma$ and $\omega$ expectation
values as a function of $\rhoB$.
We find that the $\sigma$ expectation value decreases to 
around $\sigma \simeq 60$ MeV at $\rhoB \simeq \rho_0$ in the SCL and
TM1 models, while the decrease is smaller in the BL and Boguta models. 

The modifications of $\sigma$ and $\omega$ from the vacuum values
determine the nucleon scalar and vector(-isoscalar) potentials,
$U_s(\rhoB)=-g_\sigma(f_\pi-\sigma)$ and $U_v(\rhoB)=g_\omega\omega$.
The bulk properties of nuclei such as the binding energies
and nuclear radii, which TM models have been modeled to explain,
would be mainly determined by the nuclear matter properties
and meson expectation values at around and below $\rho_0$.
Thus the above similarity of the meson fields at low densities
together with the EOS similarity
suggests that the bulk properties of finite nuclei
would be also well described in the SCL model as in TM models.

\subsection{Finite Nuclei}
\label{Subec:Finite}

In describing finite nuclei,
it is numerically preferable to represent the Lagrangian in the shifted field
$\varphi \equiv f_\pi - \sigma$
and to separate the $\sigma$ mass term from the chiral potential $V_\sigma$,
\begin{gather}
V_\sigma(\sigma)
= \frac12 m_\sigma \varphi^2 + V_\varphi(\varphi)
\ ,\quad
V_\varphi(\varphi)
=-2a_\sigma f_\SCL\left(\frac{\varphi}{f_\pi}\right)
\\
f_\SCL(x)
= \log\sbrace{1-x}
 + x + \frac{x^2}{2}
\ ,
\end{gather}
since the boundary condition is given as $\varphi \to 0$ at $r\to\infty$.
In addition, it is necessary to include the photon field
which represents the Coulomb potential.
Here we take the static and mean-field approximation for boson fields,
then RMF Lagrangian can be written as follows,
\begin{align}
\Lag_\chi^{\scriptscriptstyle \mathrm{RMF}}
&=\psibar_N\left[
	i\Slash{\partial} - M_N^\ast(\varphi) - \gamma^0 U_v(\omega,R,A)
  \right]\psi_N \nonumber\\
&- \frac{1}{2}\left(\mib{\nabla}\varphi\right)^2
                    - \frac{1}{2}m_\sigma^2 \varphi^2 - V_\varphi(\varphi)
\nonumber\\
&+ \frac{1}{2}\left(\mib{\nabla}\omega\right)^2 + \frac{1}{2}m_\omega^2 \omega^2
 + \frac{D_{\omega}}{4} \omega^4
 + \frac{1}{2}\left(\mib{\nabla} R\right)^2 + \frac{1}{2}m_\rho^2 R^2
 + \frac{1}{2}\left(\mib{\nabla} A\right)^2\ ,\label{RMFlag-f}\\
M_N^\ast(\varphi)
&=M_N-g_{\sigma}\varphi
\ ,
\quad
U_v(\omega,R,A)
 =g_{\omega}{\omega} + g_{\rho}\tau_3 R
 + e\frac{1+\tau_3}{2} A 
\ .
\end{align}
The field equations of motion derived from this Lagrangian read,
\begin{gather}
\bbrace{-i\mib{\alpha}\cdot\mib{\nabla} + \beta M^\ast + U_v}\psi_N
        = \varepsilon_i \psi_N
\ ,\label{Lag-baryon}\\
\sbrace{-\triangle + m_\sigma^2}\varphi
=  g_\sigma\rhoS - \diff{V_\varphi}{\varphi}
\ ,\label{Lag-sigma}\\
\sbrace{-\triangle + m_\omega^2}\omega
= g_\omega\rhoB
  - c_\omega \omega^3
\ ,\label{Lag-omega}\\
\sbrace{-\triangle + m_\rho^2}R = g_\rho\rhoT
\ ,\label{Lag-rho}\\
-\triangle A =  e\rhoBp
\ ,\label{Lag-photon}
\end{gather}
where $\rhoS=\rhoSp+\rhoSn$, $\rhoB=\rhoBp+\rhoBn$, $\rhoT=\rhoBp-\rhoBn$
denote scalar, baryon and isospin densities of nucleons, respectively.
Total energy is given by the integral of the energy density given as,
\begin{align}
E =& \sum_{i,\kappa,\alpha}2|\kappa|\varepsilon_{i\kappa\alpha}
- \frac{1}{2}\int \mbrace{ g_{\sigma}\varphi\rhoS + g_{\omega}\omega\rhoB
+ g_{\rho}R\rhoT + e^2A\rhoBp}d\mib{r}\nonumber\\
&+\int\sbrace{V_\varphi
- \frac12\varphi\diff{V_\varphi}{\varphi}
+ \frac{D_\omega}{4} \omega^4
} d\mib{r}
\label{totalBE}
\end{align}
where we use the Eq.(\ref{Lag-baryon})-(\ref{Lag-photon})
to calculate second order derivatives
of meson fields.
We solve the self-consistent coupled equations
(\ref{Lag-baryon})-(\ref{Lag-photon})
by iteration until the convergence of total energy is achieved.
In this work,
we assume that the nucleus which we treat is spherical,
then the nucleon wave functions are expanded
in spherical harmonic basis as follows,
\begin{gather}
        \psi_{\alpha i\kappa m}= \sbrace{
        \begin{array}{c}
                i[G^\alpha_{i\kappa}/r]\Phi_{\kappa m}\\
                -[F^\alpha_{i\kappa}/r]\Phi_{-\kappa m}
        \end{array}
        }
        \zeta_\alpha
\ ,\\
        \rhoBt =
            \sum_i^{\mbox{occ.}}\sbrace{\frac{2j_i + 1}{4\pi r^2}}
                                \sbrace{|G_i^\alpha (r)|^2 + |F_i^\alpha (r)|^2}
\ ,\\
        \rhoSt =
            \sum_i^{\mbox{occ.}}\sbrace{\frac{2j_i + 1}{4\pi r^2}}
                                \sbrace{|G_i^\alpha (r)|^2 - |F_i^\alpha (r)|^2}
\ .
\end{gather}
where $\zeta_\alpha$ represents the isospin wave function, proton or neutron,
and $\kappa = l$ ($-(l + 1)$) for $l = j-1/2$ ($l = j+1/2$).

In comparing the calculated results in mean-field models
with the experimental binding energies and charge radii, 
we have to take account of several corrections.
In this work, we consider
the center-of-mass (CM) kinetic energy correction on the total energy
and CM and nucleon size correction on nuclear charge rms radius
in the same way as that adopted in Ref.~\citen{TM1}.
The CM kinetic energy is assumed to be
\begin{equation}
        E_\mathrm{ZPE}
        = \frac{\expv{\bold{P}_\mathrm{CM}^2}}{2AM_N}
	\simeq \frac{3}{4}\hbar\omega
\ ,
\label{come}
\end{equation}
where $\bold{P}_\mathrm{CM}=\sum_i \bold{p}_i$ is the CM momentum.
This correction gives an exact result
when the state is represented by a harmonic-oscillator wave function,
and we assume that it also applies to the RMF wave functions.
The CM correction on the proton rms radius 
is written as
\begin{align}
\delta \expv{r_\mathrm{p}^2}
&=- 2\expv{\bold{R}_\mathrm{CM}\cdot\bold{R}_\mathrm{p}}
  +            \expv{\bold{R}_\mathrm{CM}^2}
\nonumber\\
&\simeq	\begin{cases}
	{\displaystyle - {3\hbar\over2AM_N\omega}}
		 	& \mbox{(for heavy nuclei)\ ,}\\[0.5ex]
       {\displaystyle
       		- \frac{2Z}{A}\expv{\bold{R}_\mathrm{p}^2}
		+ \expv{\bold{R}_\mathrm{CM}^2}
		=
		- \frac{2\expv{r_\mathrm{p}^2}}{A}
		+ \frac{\expv{r_\mathrm{M}^2}}{A}
       }
			& \mbox{(for light nuclei)\ ,}
	\end{cases}
\label{comc}
\end{align}
where $\bold{R}_\mathrm{p}=\sum_{i\in p} \bold{r}_i/Z$ is
the proton CM position,
and $\expv{r_\mathrm{p}^2}$ and $\expv{r_\mathrm{m}^2}$ represent
the proton and matter mean square radii, respectively.
We assume again that harmonic-oscillator results applies for heavy nuclei.
For light nuclei, we evaluate the correction
in RMF wave functions, and we consider only the direct-term contributions.
The charge rms radius is obtained
by including the finite size effects of protons and neutrons,
\begin{equation}
        \expv{r_\mathrm{ch}^2}
        = \expv{r_\mathrm{p}^2}
		+ \expv{r_\mathrm{ch}^2}_\mathrm{p}
		+ {N\over Z} \expv{r_\mathrm{ch}^2}_\mathrm{n}
\ ,
        \label{crms}
\end{equation}
We evaluate the binding energies and charge rms radii with these corrections,
and the pairing energy for open-shell nuclei are neglected.

\begin{table}[t]
\centering
\begin{tabular}{cccc}
\multicolumn{2}{l}{Parameters}\\
\toprule
 $m_\sigma$ (MeV)	& $g_\omega$	& $g_\rho$	& $c_\omega$\\
\midrule
 502.63		& 13.02		& 4.40		& 200\\
\bottomrule
\end{tabular}
\begin{tabular}{cccccc}
\multicolumn{2}{l}{Constants}\\
\toprule
 $M_N$ (MeV) & $f_\pi$ (MeV) & $m_\pi$ (MeV) & $m_\omega$ (MeV) & $m_\rho$ (MeV)
 & $g_\sigma=M_N/f_\pi$\\
\midrule
 938 &  93 & 138 & 783 & 770  & 10.08 \\
\bottomrule
\multicolumn{6}{c}{} \\
\toprule
   $\hbar c$ (MeV$\cdot$fm)
 & $\rho_0$ ($\mbox{fm}^{-3}$) & $E/A(\rho_0)$ (MeV)
 & $\hbar\omega$ (MeV) 
 & $ \expv{r_\mathrm{ch}^2}_p$
 & $ \expv{r_\mathrm{ch}^2}_n$\\
\midrule
   197.32705 & 0.145 & $-16.3$ & $41\, A^{-1/3}$
 & $(\mbox{0.862 fm})^2$ & $-(\mbox{0.336 fm})^2$ \\
\bottomrule
\end{tabular}
\caption{
Parameters and constants adopted in the present work.
Two of the parameters ($g_\omega$ and $c_\omega$) are determined
to fit the saturation point of symmetric nuclear matter
and others ($m_\sigma$ and $g_\rho$) are fixed through global fitting
of Sn and Pb isotopes' binding energies.
} 
\label{Table:paramset}
\end{table}

In describing finite nuclear properties,
we have two free parameters, $g_\rho$ and $m_\sigma$,
which we cannot determine from
the symmetric nuclear matter saturation point.
We have fitted the binding energies of Sn $(Z=50)$ and Pb $(Z=82)$ isotopes and
have fixed the parameter values as $g_\rho=4.40$ and $m_\sigma=502.63$ MeV.
Other parameters are obtained to fit the symmetric nuclear matter
saturation point, and summarized in Table \ref{Table:paramset}.
By using this parameter set,
we have calculated the binding energies and charge rms radii
of C, O, Si, Ca, Ni, Zr, Sn and Pb isotopes.

\begin{table}[tb]
\centering
\begin{tabular}{cccccccccc}
        \toprule
        \multicolumn{10}{c}{$E/A$ (MeV)} \\
        \midrule
          Nucleus      & exp. & SCL  & TM1  & TM2  & NL1  & NL3  & I/110 
         & IF/110 & VIIIF/100 \\
        \midrule                                                         
          ${}^{12}$C   & 7.68 & 7.09 &   -  & 7.68 &   -  & -    & -     
         &   -    & -         \\
          ${}^{16}$O   & 7.98 & 8.06 &   -  & 7.92 & 7.95 & 8.05 & 7.35  
         &  7.86  & 7.18      \\
          ${}^{28}$Si  & 8.45 & 8.02 &   -  & 8.47 & 8.25 & -    & -     
         &  -     & -         \\
          ${}^{40}$Ca  & 8.55 & 8.57 & 8.62 & 8.48 & 8.56 & 8.55 & 7.96  
         &  8.35  & 7.91      \\
          ${}^{48}$Ca  & 8.67 & 8.62 & 8.65 & 8.70 & 8.60 & 8.65 & -     
         &  -     & -         \\
          ${}^{58}$Ni  & 8.73 & 8.54 & 8.64 &   -  & 8.70 & 8.68 & -     
         &  -     & -         \\
          ${}^{90}$Zr  & 8.71 & 8.69 & 8.71 &   -  & 8.71 & 8.70 & -     
         &  -     & -         \\
          ${}^{116}$Sn & 8.52 & 8.51 & 8.53 &   -  & 8.52 & 8.51 & -     
         &  -     & -         \\
          ${}^{196}$Pb & 7.87 & 7.87 & 7.87 &   -  & 7.89 & -    & -     
         &  -     & -         \\
          ${}^{208}$Pb & 7.87 & 7.87 & 7.87 &   -  & 7.89 & 7.88 & 7.33  
         &  7.54  & 7.44      \\
        \bottomrule
        \multicolumn{10}{c}{} \\
        \toprule
        \multicolumn{10}{c}{charge rms radius (fm)} \\
        \midrule
          Nucleus      & exp. & SCL  & TM1  & TM2  & NL1  & NL3  & I/110 
         & IF/110 & VIIIF/110 \\
        \midrule
          ${}^{12}$C   & 2.46 & 2.43 &   -  & 2.39 &   -  & -    &  -    
         &  -     & -         \\
          ${}^{16}$O   & 2.74 & 2.62 &   -  & 2.67 & 2.74 & 2.73 &  2.64 
         &  2.62  & 2.69      \\
          ${}^{28}$Si  & 3.09 & 3.04 &   -  & 3.07 & 3.03 & -    &  -    
         &  -     & -         \\
          ${}^{40}$Ca  & 3.45 & 3.44 & 3.44 & 3.50 & 3.48 & 3.47 &  3.41 
         &  3.40  & 3.45      \\
          ${}^{48}$Ca  & 3.45 & 3.46 & 3.45 & 3.50 & 3.44 & 3.47 &  -    
         &  -     & -         \\
          ${}^{58}$Ni  & 3.77 & 3.77 & 3.76 &   -  & 3.73 & 3.74 &  -    
         &  -     & -         \\
          ${}^{90}$Zr  & 4.26 & 4.27 & 4.27 &   -  & 4.27 & 4.29 &  -    
         &  -     & -         \\
          ${}^{116}$Sn & 4.63 & 4.62 & 4.61 &   -  & 4.61 & 4.61 &  -    
         &  -     & -         \\
          ${}^{196}$Pb &   -  & 5.48 & 5.47 &   -  & 5.47 & -    &  -    
         &  -     & -         \\
          ${}^{208}$Pb & 5.50 & 5.54 & 5.53 &   -  & 5.57 & 5.58 &  5.49 
         &  5.49  & 5.53      \\
        \bottomrule
\end{tabular}
\caption{Experimental and calculated binding energy and
         charge rms radius values of stable nuclei.
         Calculated results in the SCL model are compared
	 with those in
	 TM1~\protect{\cite{TM1}},
	 TM2~\protect{\cite{TM1}},
	 NL1~\protect{\cite{NL1}},
	 NL3~\protect{\cite{NL3}},
         glueball model (I/110)~\protect{\cite{Heide1994}},
         frozen glueball model (IF/110 and VIIIF/100)~\protect{\cite{Heide1994}}
         and experimental ones, respectively.
         }
\label{Table:beandcrms}
\end{table}
\begin{figure}[tbh]
        \centering
		\PSfig{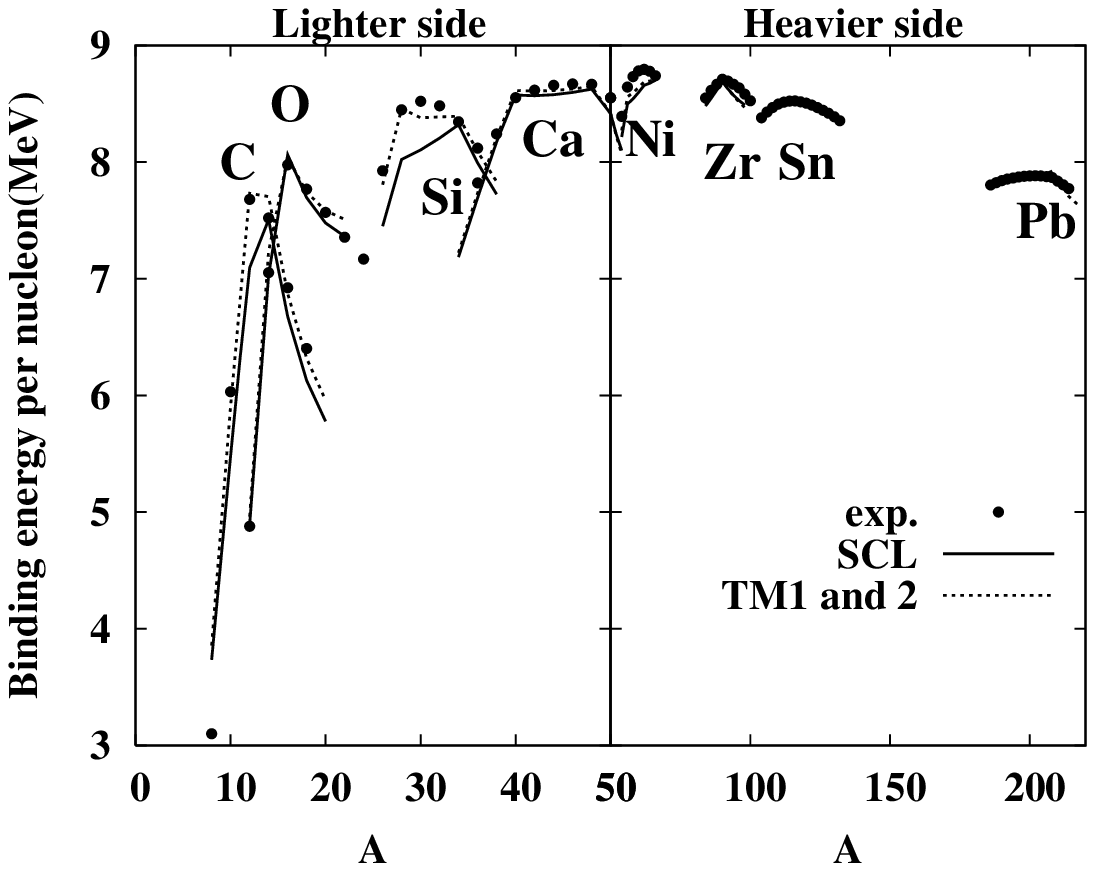}{340pt,clip}
        \caption{
Nuclear binding energies per nucleon of $Z$ (sub-)closed nuclei.
Calculated results in the SCL model (solid lines)
are compared with the experimental data (points)
and the calculated results in the TM1\protect{\cite{TM1}} model (dotted line).
}
        \label{Fig:bepn}
\end{figure}
In Table~\ref{Table:beandcrms},
we show the calculated results of
binding energies per nucleon and charge rms radii
of doubly (sub-s)closed stable nuclei,
\nuc{12}C, \nuc{16}O, \nuc{28}{Si}, \nuc{40}{Ca}, \nuc{48}{Ca},
\nuc{58}{Ni}, \nuc{90}{Zr}, \nuc{116}{Sn}, \nuc{206}{Pb} and \nuc{208}{Pb},
in comparison with the experimental data and the results
in TM~\cite{TM1}, NL1~\cite{NL1} and NL3~\cite{NL3} models.
We find good overall agreement of the SCL results
with the experimental data for heavy nuclei.
For light nuclei,
we underestimate the binding energies of nuclei,
\nuc{12}{C}, \nuc{28}{Si} and \nuc{58}{Ni}, by 0.2$-$0.6 MeV.
These nuclei have proton or neutron numbers of $Z(\mbox{or}~N)=$6, 14 and 28,
{\em i.e.}, they are $jj$ closed nuclei.
These underestimates imply that the spin-orbit interaction in the SCL model
is not enough to explain the splitting.
There are still discussions on the strength of the spin-orbit interactions
in RMF~\cite{Bhagwat},
and it is recently suggested that the explicit role of pions has large effects
in $jj$ closed nuclei~\cite{IST04,ION05,Bouyssy}.
Thus the underestimate may be related to the explicit pion effects,
appeared clearly with the restriction from the chiral symmetry.

\begin{figure}[tbh]
  \centerline{\PSfig{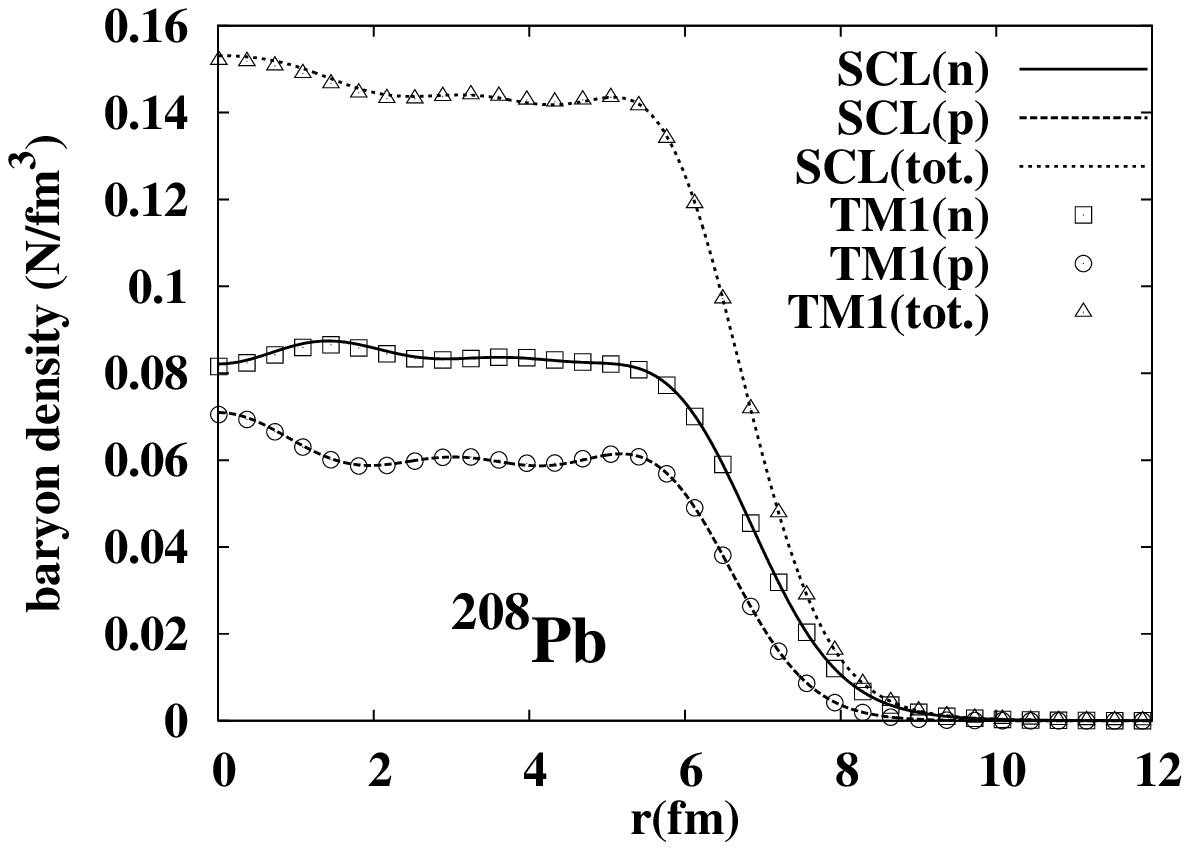}{10cm}}
  \centerline{\PSfig{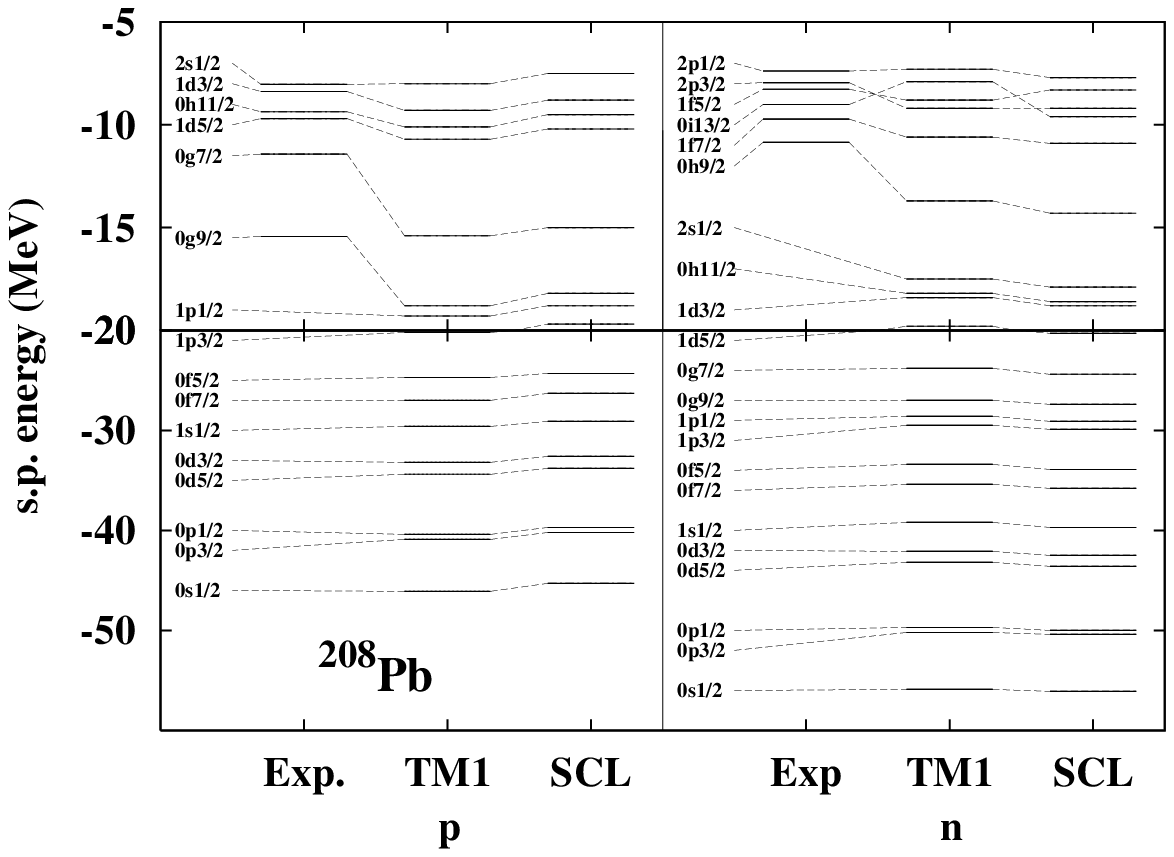}{12cm}}
  \caption{
  Comparison of nuclear density (top) and single-particle energies
  in the SCL and TM1~\cite{TM1} models for \nuc{208}{Pb} nucleus (bottom).
  Calculated results in SCL model are compared with those in TM1 model
  ~\cite{TM1} and experimental data in Ref.~\citen{VB}.
  }
  \label{Fig:Pb_rhos}
\end{figure}

In Fig.~\ref{Fig:bepn}, we show the binding energies per nucleon
in C, O, Si, Ca, Ni, Zr, Sn and Pb isotopes.
Except for the above mentioned underestimate of the spin-orbit splittings
and the underestimates in heavy Zr isotopes
which would be due to the deformation~\cite{TM1,Hirata},
binding energies are well explained in one parameter set
in the present SCL model from C to Pb isotopes.
In Fig.~\ref{Fig:bepn},
we also find that the calculated results of binding energies in the SCL model
for heavy nuclei are very close to those in the TM1 model.
Here we compare the SCL model results with those in the TM1 model
without pairing corrections.
As shown in the top panel of Fig.~\ref{Fig:Pb_rhos},
these two models give similar results of nuclear densities.
The agreement in the density distributions
is not surprising since the behavior of
$\sigma$ and $\omega$ mesons as well as the EOS
is very similar in these two models at low densities,
as shown in Fig.~\ref{Fig:EOSes-UsvComp}.
In the bottom panel of Fig.~\ref{Fig:Pb_rhos}, 
we find some differences and changes of the order
in the single particle energies,
especially in those for neutrons around the Fermi energy,
which would be affected by the model details.

There are several other chirally symmetric models
which well describes the nuclear matter as well as 
finite nuclei of \nuc{16}{O}, \nuc{40}{Ca} and
\nuc{208}{Pb}~\cite{Heide1994,Furnstahl1995}.
In these models, they introduce the broken scale invariance
through the glueball field,
and this glueball couples to $\sigma$ in a logarithmic term,
which is conjectured under the requirement of the scale invariance.
In Ref.~\citen{Heide1994},
Heide et al. proposed a chiral effective Lagrangian containing the glueball,
and examined that model through nuclear matter and finite nuclear studies.
With some of their parameter sets (I/110, IF/110),
one can describe the finite nuclear properties reasonably well,
but these parameters give stiff nuclear matter EOS, $K>340$ MeV.
With the parameter set of VIIIF,
one can obtain reasonably soft EOS ($K=267$ MeV),
but the binding energies of nuclei are underestimated by $0.4-0.8$ MeV
per nucleon for \nuc{16}O, \nuc{40}{Ca} and \nuc{208}{Pb}.
In Ref.~\citen{Furnstahl1995},
Furnstahl et al. extended the terms of the chiral effective Lagrangian
with glueball having eight free parameters.
In their work, the binding energies of
\nuc{16}O, \nuc{40}{Ca} and \nuc{208}{Pb} nuclei are well described
(up to 0.4 MeV per nucleon deviation from the experimental data),
and reasonably soft EOS's ($K=(194-244)$ MeV) are obtained.
Compared to these models, the present SCL model has only four free parameters
which have been determined to fit the symmetric nuclear matter saturation point
and finite nuclear properties.
The main difference in the SCL model from these two glueball models
is in the $\omega$ self-interaction term,
$c_\omega \omega^4$, which simulates the suppression of $\omega$
at large densities,
which has been demonstrated to emerge
in the Dirac-Br\"{u}ckner HF theory.\cite{RBHF}

\section{Summary}
\label{Sec:Summary}
In this work, we have developed
a relativistic mean field (RMF) model
with a chiral symmetric effective potential $V_\sigma$
(vacuum energy density as a function of $\sigma$)
having a logarithmic term
derived in the strong coupling limit (SCL) of the lattice QCD.
The logarithmic potential term of $\sigma$
is found to have favorable features;
it prevents the normal vacuum from collapsing at low densities,
and it does not have any instabilities as a function of $\sigma$.
By introducing vector mesons ($\omega$ and $\rho$)
and a non-linear vector meson self-interaction term
($(\omega_\mu\omega^\mu)^2$)
in a phenomenological way to fit the nuclear matter saturation point
and the binding energies of finite nuclei,
we have demonstrated that both of
symmetric nuclear matter equation of state (EOS)
and finite nuclear properties are well described
in one parameter set containing four free parameters.
The obtained EOS is comparable 
to those in phenomenologically successful but non-chiral models
such as TM1,\cite{TM1} 
and to that in a chiral model containing phenomenologically added
higher order terms of $\sigma$.\cite{SO2000}
Binding energies of finite nuclei are also well reproduced in a wide mass range
from C to Pb isotopes, while the binding energies of several $jj$ closed nuclei 
are underestimated, suggesting smaller spin-orbit interactions
in the present model.

Comparisons of the present SCL model are made with other chiral RMF models
at vacuum and finite densities in symmetric nuclear matter.
We have demonstrated that other chiral RMF models
have some problems in simultaneous description both of nuclear matter
and finite nuclei:
The baryon loop (BL)~\cite{BL} and Sahu-Ohnishi (SO)~\cite{SO2000}
models have instability at large $\sigma$ values,
and the linear $\sigma$ model with the $\sigma^2\omega^2$ coupling
(Boguta model) gives too stiff EOS of symmetric nuclear matter.
While the chiral RMF models with glueballs,
which represent the broken scale invariance,
give reasonably good description both of nuclear matter and finite nuclei, 
introducing unobserved glueballs may not be justified in hadronic models.

The energy functional in the present SCL model
seems to be very similar to that in the TM1 model at low densities.
This point has been examined in the symmetric nuclear matter EOS
and finite nuclear properties such as
binding energies, charge rms radii, nuclear densities
and single particle levels.
At high densities, the SCL model gives a little softer EOS than that in TM1.
This softness has some effect on neutron star properties
and supernova explosion energies, which will be reported elsewhere.

There are several directions which should be investigated further.
First, the binding energy underestimate problem in $jj$ closed nuclei
should be studied in an extended framework
including explicit role of pions~\cite{IST04,ION05}.
Next, it would be desirable to extend the present model to that
with the flavor SU(3) chiral symmetry.
Preliminary works in this direction have been already made,
which suggest that we can obtain a softer symmetric nuclear EOS
($K \simeq 210$ MeV) due to the coupling to hidden strangeness.
Thirdly, the finite temperature effects have not been included
in the effective potential $V_\sigma$ adopted in this work.
Since the present $V_\sigma$ diverges at $\sigma \to 0$,
it cannot describe the chiral phase transition at high temperatures.
There are several works which include finite temperature effects
in the strong coupling limit of the lattice QCD,\cite{Nishida2004,KMOO06}
and these works have shown that the effective potential
is smoothed at finite temperatures,
and it is possible to describe chiral phase transitions.

\section*{Acknowledgements}
We would like to thank Dr. K. Naito for important discussions
in the initial stage of this work.
This work is supported in part by the Ministry of Education,
Science, Sports and Culture, Grant-in-Aid for Scientific Research
under the grant numbers,
    15540243		
and 1707005.		

%

\end{document}